\documentclass[%
 reprint,
 amsmath,amssymb,
 aip,
 jcp,
]{revtex4-1}

\usepackage{amsmath}
\usepackage{graphicx}
\usepackage{dcolumn}
\usepackage{bm}
\usepackage{epsfig,color,xspace,multirow,xr,bbold}
\usepackage[all]{xy}
\usepackage{setspace}
\usepackage{url}

\usepackage{booktabs} 

\usepackage{enumitem} 
\setlist[itemize]{noitemsep} 

\usepackage{hyperref} 

\usepackage{epsfig}
\usepackage{overpic}
\usepackage{tikz}
\usepackage{amsmath}
\usepackage{pifont}

\begin{document}


\title{Polymorphism of syndiotactic polystyrene crystals from
  multiscale simulations}
\author{Chan Liu}
\author{Kurt Kremer}
\author{Tristan Bereau}
\email{bereau@mpip-mainz.mpg.de}

\affiliation{Max Planck Institute for Polymer Research, Ackermannweg
  10, 55128 Mainz, Germany}

\date{\today} 
\begin{abstract}
  Syndiotactic polystyrene (sPS) exhibits complex polymorphic behavior
  upon crystallization. Computational modeling of polymer
  crystallization has remained a challenging task because the relevant
  processes are slow on the molecular time scale.  We report herein a
  detailed characterization of sPS-crystal polymorphism by means of
  coarse-grained (CG) and atomistic (AA) modeling.  The CG model,
  parametrized in the melt, shows remarkable transferability
  properties in the crystalline phase.  Not only is the transition
  temperature in good agreement with atomistic simulations, it
  stabilizes the main $\alpha$ and $\beta$ polymorphs, observed
  experimentally.  We compare in detail the propensity of polymorphs
  at the CG and AA level and discuss finite-size as well as
  box-geometry effects.  All in all, we demontrate the appeal of CG
  modeling to efficiently characterize polymer-crystal polymorphism at
  large scale.
\end{abstract}


\maketitle


\section{Introduction}

The crystallization of molecular assemblies often involves the
stabilization of a variety of distinct conformers, denoted
polymorphism.\cite{Bernstein2002} The impact of polymorphs, may it be
thermodynamic, structural, or physiological, makes their understanding
and prediction essential for a broad area of research, from hard
condensed matter to organic materials.  The computational modeling of
crystalline materials has rapidly developed in recent years, from
crystallization~\cite{Perego2015, Radu2017} to predicting polymorphs
and their relative stability~\cite{Beran2016, Price2014, Marom2013,
  Reilly2014}.  These developments are remarkable given the complexity
of the system: from the molecular architecture to sampling challenges
at low temperatures to the significant timescales involved.  For this
reason, past computational studies focused on crystallization and/or
polymorphism have been limited to molecular crystals---based on small
molecules.  Here instead, we extend the computational characterization
of polymers crystals,\cite{Hu2005, Sommer2010, Yamamoto2013,
  Welch2017} by modeling for the first time its self assembly and
resulting polymorphs.

\begin{figure}
\includegraphics[scale=0.6]{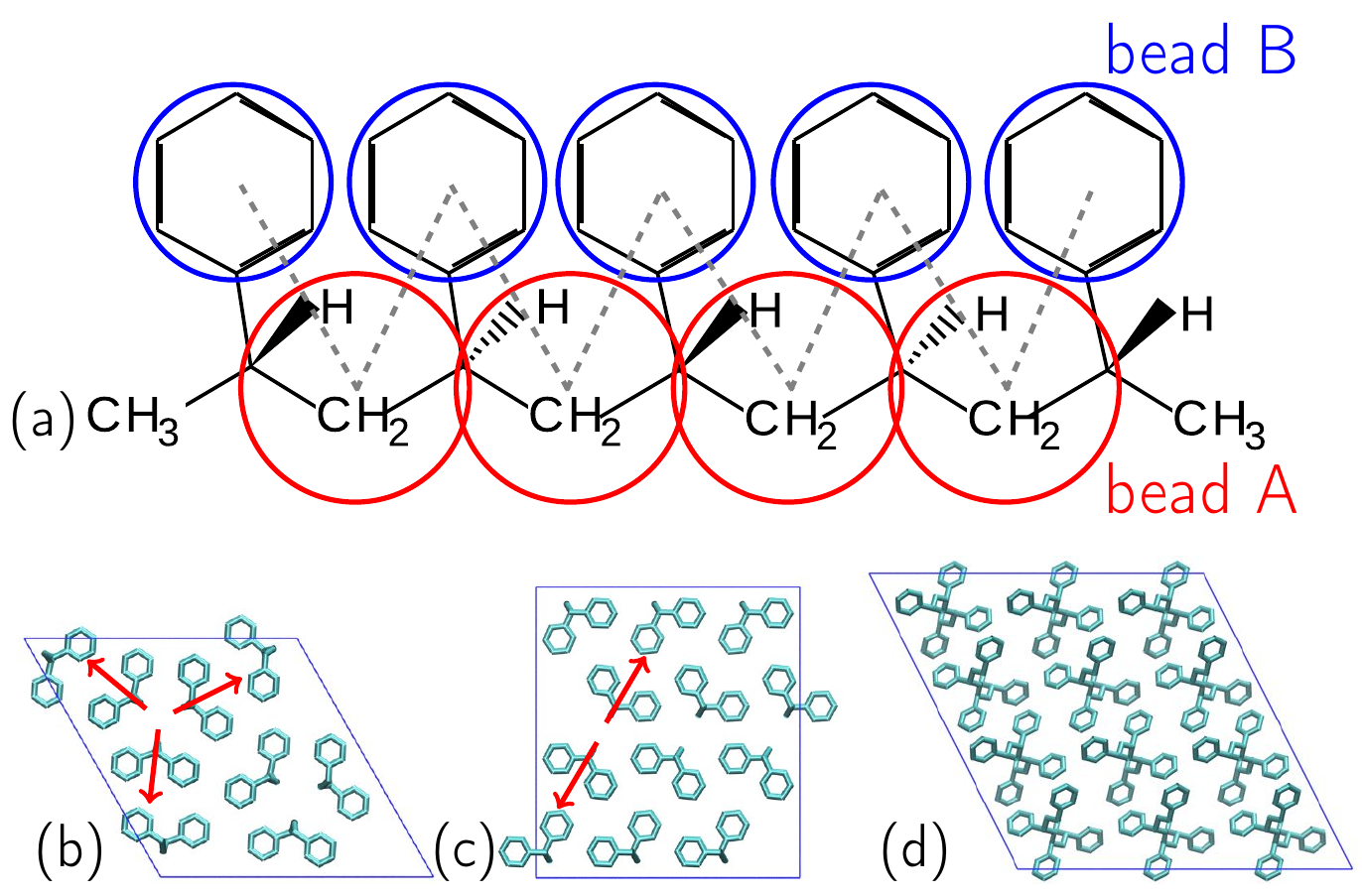}
\caption{(a) Representation of syndiotactic polystyrene, including the
  CG mapping scheme: each monomer is mapped onto two coarse-grained
  beads. CG bead A is the center of mass of the backbone connecting
  two sidechains, while bead B is the center of mass of the phenyl
  group. Polymorphism: (b) $\alpha$, (c) $\beta$, (d) $\delta$ forms.}
\label{fig:mapping}
\end{figure}

Syndiotactic polystyrene (sPS) is known for its unusual crystal
polymorphism. Five different crystalline forms have been reported
experimentally.  Fig.~\ref{fig:mapping} shows three main forms studied
in this work.  The unit cells are projected onto the $a-b$ plane, and
the backbones of the chains are along the $c$ axis.  Experimentally,
two types of crystalline phases have been identified: the
zig-zag-chain forming
$\alpha$~\cite{Rosa1991,Corradini1994,Rosa1996,Cartier1998} and
$\beta$~\cite{Rosa1992,Chatani1993}
appear upon thermal annealing of a melt, whereas the other three
helix-forming crystalline phases, $\gamma$~\cite{Rizzo2002,Rizzo2005},
$\delta$~\cite{Rosa1997,Milano2001,Gowd2006} and
$\varepsilon$~\cite{Rizzo2007,Petraccone2008,Tarallo2010}
are obtained by solution processing. The $\alpha$ and $\beta$ forms of
sPS are further classified into the limiting disordered forms
($\alpha'$ and $\beta'$) and limiting ordered forms ($\alpha''$ and
$\beta''$)~\cite{Guerra1990,Rosa1991,Rosa1996,Rosa1992}. In
particular, melt crystallization procedures generally produce the
limiting ordered $\alpha''$ and limiting disordered $\beta'$
models~\cite{Guerra1990}. The limiting disordered $\alpha'$ model is
obtained by annealing the amorphous sample~\cite{Guerra1990}, whereas
the limiting ordered $\beta''$ model is obtained by crystallization
from solution, when the solvent is rapidly removed at higher
temperatures above 150$^{\circ}C$~\cite{Rosa1992}. Two of the helical
crystalline phases ($\delta$ and $\varepsilon$) can only be obtained
by guest removal from co-crystalline phases. The $\delta_e$ form is
transformed into the solvent-free $\gamma$ form by annealing above
130$^{\circ}C$~\cite{Guerra1990,Rosa1997,Rizzo2002}.  These findings
illustrate that the experimentally-observed structures depend on the
experimental processing, making conclusions about their thermodynamic
equilibrium difficult.

Several molecular simulation studies have been performed to better
understand the nanoporous cavity structures formed by crystalline
syndiotactic polystyrene. Tamai and his co-workers~\cite{Tamai2003}
studied the size, shape and connectivity of the cavities in the
crystal $\alpha$, $\beta$ and $\delta$ forms.  Some other properties
were also studied, such as diffusion of gases~\cite{Milano2002,
  Tamai2003-1}, reorientational motion of guest
solvents~\cite{Tamai2003-2, Tamai2005}, and sorption of small
molecules~\cite{Figueroa-Gerstenmaier2010,
  Figueroa-Gerstenmaier2010-1, Sanguigno2011}.  While these studies
helped understand the behavior of specific forms of sPS crystals, they
did not address the crystallization process or the relative stability
of the polymorphs. By studying sPS on the nanosecond timescale, they
could not observe self assembly and spontaneous polymorph
interconversion, due to their metastability.  Yamamoto observed the
onset of isotactic polypropylene crystallization, but leading to a
smectic mesophase instead of crystal
polymorphs~\cite{Yamamoto2014}. The simulation of polymer
crystallization remains challenging because these processes are slow
on the molecular time scale.  In addition, the simulated chain lengths
must be large enough to link to experimentally realistic
situations~\cite{strobl2007}.  As an alternative to atomistic (AA)
simulations, coarse-grained (CG) models~\cite{Mueller-Plathe2002,
  Voth2008, Peter2010} combine a number of atoms into superatoms or
beads to significantly speed up the simulations, while potentially
retaining enough chemical detail to differentiate polymorphs.  Its
capability to reach longer length and timescales is strengthened by an
observation we present below: the relatively small systems we can
afford at the AA level display finite-size effects.

Several CG models for polystyrene have been reported, but so far
focusing solely on the amorphous phase.  In several models one CG bead
represents a single monomer~\cite{Milano2005, Spyriouni2007, Qian2008,
  Sun2006}.  Milano and Mueller-Plathe~\cite{Milano2005,
  Spyriouni2007} center the beads on methylene carbons.  The model of
Qian and coworkers~\cite{Qian2008} places the bead on the center of
mass of the monomer.  Both of these models use different bead types to
represent different types of diads, thus keeping information about the
chain stereosequences.  Sun and Faller~\cite{Sun2006} center the beads
on the backbone carbons to which the phenyl rings are attached. In
this model, each bead represents one PS monomer without distinguishing
between different types of diads.  Single-bead representations for a
polystyrene monomer represent significant issues to describe
polymorphism, as they lack the necessary degrees of freedom to
distinguish between the relevant forms.  Models of higher resolution
were also developed: Harmandaris and coworkers~\cite{Harmandaris2006,
  Harmandaris2007} devised two different two-bead-per-monomer
models. In their first model~\cite{Harmandaris2006}, the CH$_2$ group
of the backbone chain is represented by one CG bead, whereas the
remaining CH group of the monomer in the backbone and its pendant
phenyl ring are represented by another CG bead. In their second
model~\cite{Harmandaris2007}, the phenyl ring is represented by one CG
bead, while the other bead represents the CH$_2$ of the backbone as
well as a contribution from each one of the two neighboring CH groups.
By comparison, the second mapping scheme better reproduces the local
chain conformations and melt packing observed in atomistic
simulations. Based on this mapping scheme, Fritz and
coworkers~\cite{Fritz2009} developed a new CG force field using the
conditional reversible work (CRW) method~\cite{Brini2011,
  Deichmann2017}, a thermodynamic-based CG method that samples
isolated atomistic chains or pairs of oligomers in vacuum.  In this
paper, we find that this CG model can successfully crystallize sPS.
The bonded force field seems to adequately reproduce the local chain
conformations of syndiotactic polystyrene so that the sterics allow
for crystallization.  More importantly, this finding illustrates the
remarkable transferability properties of the CRW method.

In this paper, we report both AA and CG simulations of the
temperature-driven phase transition between crystal and melt.
Annealing simulations led to highly-ordered conformations, but
exhibiting significant hysteretic behavior.  Replica-exchange
molecular dynamics (REMD) simulations helped the conformational
sampling and the identification of the transition temperature.
Backmapping of CG snapshots provided starting AA configurations.  For
both resolutions, we consider several polymorphs, $\alpha$, $\beta$
and $\delta$, observed experimentally. We find that the two models
stabilize $\alpha$ and $\beta$, but not $\delta$.  We discuss the
impact of finite-size effects and box geometries.  To alleviate these
artefacts, we use the CG model to construct a larger system and
discuss the convergence of polymorph populations.


\section{Methods}

\subsection{Model}

\subsubsection{AA Model}

We used the atomistic (AA) model of polystyrene from
Mueller-Plathe~\cite{Mueller-Plathe1996}.  In this model, the phenyl
groups are described using parameters from benzene, while the
parameters for the aliphatic carbons and hydrogens were identical to
aliphatic polymers.  The rotation of the phenyl rings is left free.
All bond lengths were constrained by the LINCS method~\cite{Hess1997}.

\subsubsection{CG Model}

The CG polystyrene force field of Fritz \emph{et al.}~\cite{Fritz2009}
is derived from the above-mentioned AA model. It maps each monomer
onto two CG beads of different types, denoted A for the chain backbone
and B for the phenyl ring (see Fig.~\ref{fig:mapping}a). We stress
that the CG model represents PS by a \emph{linear} chain, where the
beads are connected by CG bonds A--B. There are no bonds between the A
beads, and the close connection between them is reproduced indirectly
by the angular potentials $\theta _{ABA}$. The tacticity of
polystyrene is encoded in the potential parameters. Here we only use
the parameters of syndiotactic polystyrene whose chains only consist
of racemic diads.

The CG force field includes the bonded and nonbonded interaction
potentials in a tabulated form and are derived separately. Potentials
for bonded degrees of freedom of the CG model are obtained by direct
Boltzmann inversion of distributions obtained from atomistic
simulations of single chains in vacuum using stochastic dynamics.

The nonbonded potentials are derived by the conditional reversible
work (CRW) method~\cite{Brini2011,Deichmann2017}. They are obtained
from constraint dynamics runs with the all-atom model of two trimers
(or fourmers) in vacuum. In these runs the atoms mapping to a pair of
beads A or B were held at fixed distance $r$. The pair potential of
mean force (PMF), $V_{\mathrm{PMF}}$, was calculated by
\begin{equation} \label{eq:PMF}
  V_{\mathrm{PMF}}(r) = \int^{r}_{r_m}\mathrm{d}s\left[\langle
    f_c\rangle_s\right]+2k_BT\ln r,
\end{equation} 
where $k_B$ is the Boltzmann constant, $T$ is the temperature, $f_c$
is the constraint force between the two CG mapping points, $\langle
\cdot \rangle_s$ is a constrained ensemble average, and $r_m$ is the
maximum distance between the two mapping points.

The effective, nonbonded A--A interaction potential is next obtained
from
\begin{equation} \label{eq:CRW}
  V^{\mathrm{AA}}_{\mathrm{eff}}(r) =
  V^{\mathrm{AA}}_{\mathrm{PMF}}(r) - V^{\mathrm{excl,
      AA}}_{\mathrm{PMF}}(r)
\end{equation}
where the second PMF, $V^{\mathrm{excl, AA}}_{\mathrm{PMF}}$, is along
the same coordinate $r$ but excludes all direct A--A atomistic
interactions while maintaining all other interactions with and between
neighboring parts of the oligomers.  A similar procedure is applied to
A--B and B--B.

One advantage of this approach is that it is computationally
inexpensive. More importantly, this approach is widely different from
many other CG methods that use information on condensed-phase
properties of the melt state as input in the development of the
effective potentials.  This crucial difference enables the greater
transferability reported herein.

\subsection{Replica exchange simulations}

Replica exchange molecular dynamics (REMD)~\cite{Sugita1999} is used
to enhance the sampling of computer simulations, especially when
metastable states are separated by relatively high energy barriers. It
involves simulating multiple replicas of the same system at different
temperatures and randomly exchanging the complete state of two
replicas at regular intervals with a Metropolis criterion.

All simulations reported in this study were performed using the
molecular dynamics package GROMACS 4.6~\cite{Hess2008}.  The
integration time step of the AA model was 1 fs. AA-RE simulations were
performed with 48 (56 for the $\beta$ form) temperatures from 250 K to
550 K (700 K for the $\beta$ form), and replica exchange was attempted
every 1 ps.  Simulations were carried out in the $NPT$ ensemble using
the stochastic velocity rescaling thermostat~\cite{Bussi2007}
($\tau_T^{AA}=0.2$ ps) and the Berendsen barostat~\cite{Berendsen1984}
($\tau_P^{AA}=2$ ps). For the CG model, the integration time step was
$10^{-3}~\tau$, where $\tau$ defines the reduced unit of time in the
CG model. CG-RE simulations were performed with 48 temperatures from
250 K to 550 K, and replica exchange was attempted every $\tau$. In CG
simulations, we also employed the stochastic velocity rescaling
thermostat ($\tau_T^{CG}=0.2~\tau$) and the Berendsen barostat
($\tau_P^{CG}=2~\tau$). Note that in all replica exchange simulations,
we use isotropic pressure coupling, which means that the angles of the
simulation box are fixed. We have found that anisotropic coupling led
to large fluctuations close to the phase transition.

\subsection{Order parameter}

\begin{figure}
\includegraphics[scale=0.7]{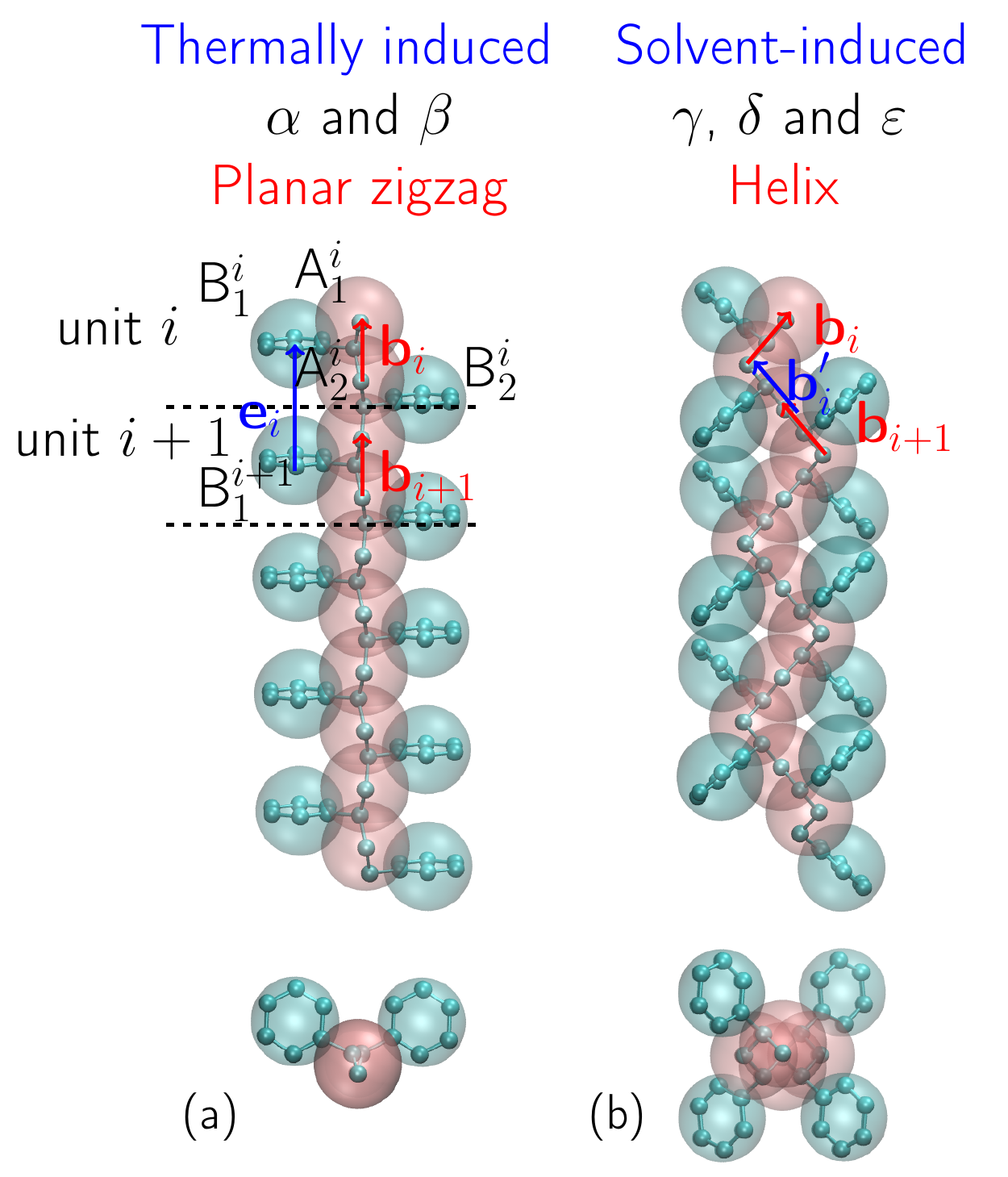}
\caption{Five different crystalline forms: two crystalline phases
  ($\alpha$ and $\beta$) with trans-planar chains; three crystalline
  phases ($\gamma$, $\delta$ and $\varepsilon$) with the s(2/1)2
  helical chains.}
\label{fig:conformation}
\end{figure}

Orientational order parameters are useful indicators of crystal
formation and crystal growth. In this work, calculating order
parameters in both AA and CG simulations is based on geometries at the
CG level, and we take one racemic diad (two monomers) as one
characterizing unit. In a polymer chain, for a side-chain bead B$_1^i$
in unit $i$ (see Fig.~\ref{fig:conformation}), we identify an
orientation vector which points from B$_1^i$ to B$_1^{i+1}$ and
normalize it to a unit vector
\begin{equation} \label{eq:ori_vector}
  {\bf e}_i = \frac{{\bf r}_{i+1} -
    {\bf r}_{i}}{\left|
    {\bf r}_{i+1}-{\bf r}_{i}\right|}.
\end{equation}
Similar orientation vectors can be obtained for bead A$_1^i$, A$_2^i$
and B$_2^i$.  For two given orientation vectors ${\bf e}_i$ and ${\bf
  e}_j$, the order parameter between them is calculated as
follows~\cite{Liu1998}:
\begin{equation} \label{eq:order_para}
  P_{ij}=\dfrac{3}{2}({\bf e}_i\cdot {\bf e}_j)^2-\dfrac{1}{2}.
\end{equation}
The order parameter of the whole system is the average over all
possible $P_{ij}$ in the system.

Note that this order parameter has some limitations: it can only
recognize trans-planar chain conformations. As mentioned in the
introduction, syndiotactic polystyrene has five different crystalline
forms. To further study the phase transition between these different
forms, we developed a procedure to characterize the different forms in
these ordered states.

\subsection{Form characterization}

In analogy to the order parameter, form characterization in both AA
and CG simulations is based on conformations at the CG level, and we
take two monomers as one characterizing unit.

There are two types of single-chain conformations (see
Fig.~\ref{fig:conformation}): trans-planar chains ($\alpha$ and
$\beta$) and helical chains ($\gamma$, $\delta$ and $\varepsilon$).
First, we identify these two different crystalline chain conformations
from amorphous chains. In a polymer chain, for unit $i$, we identify a
backbone vector ${\bf b}_i$ which points from bead A$_2^i$ to bead
A$_1^i$.  When the angle between vector ${\bf b}_i$ and vector ${\bf
  b}_{i+1}$ is around 0$^{\circ}$, the unit $i$ and $i+1$ are
trans-planar chain segments (see Fig.~\ref{fig:conformation}
(a)). When the angle between vector ${\bf b}_i$ and vector ${\bf
  b}_{i+1}$ is around 70$^{\circ}$ and the angle between vector ${\bf
  b'}_i$ (from bead A$_1^{i+1}$ to bead A$_2^i$) and vector ${\bf
  b}_i$ is around 25$^{\circ}$, the unit $i$ and $i+1$ are helical
chain segments (see Fig.~\ref{fig:conformation} (b)). The remaining
segments are considered amorphous.

For trans-planar chain segments, we discriminate crystal regions
between different chain-stretching directions. Then in each region and
each single layer, which is perpendicular to the chain stretching
direction, we measure the distances between all beads of type A in the
same layer. When the distances between two such beads is smaller than
0.6 nm, we associate them into the same group. In each group, we
measure the orientations of side chains to identify $\alpha$ and
$\beta$ forms (see Fig.~\ref{fig:mapping}(b)(c)). If the orientations
are all parallel, the group is a ``full $\beta$'' group; if the angles
of the orientations are all around 120$^{\circ}$, the group is a
``full $\alpha$'' group.  Some groups have features of both $\alpha$
and $\beta$ forms, in which case we call them ``mix" (see
Fig.~\ref{fig:CG-beta}).  In case only a single characterizing unit is
found in the group, we call it ``defect.''

In this work, we focus on the main $\alpha$ and $\beta$ polymorphs
found in annealing experiments, while all helical chain segments are
herein called $\delta$ for convenience.


\section{Results and discussion}

\subsection{Crystallization of the CG model}

\subsubsection{CG annealing simulations}

\begin{figure}
\includegraphics[scale=0.6]{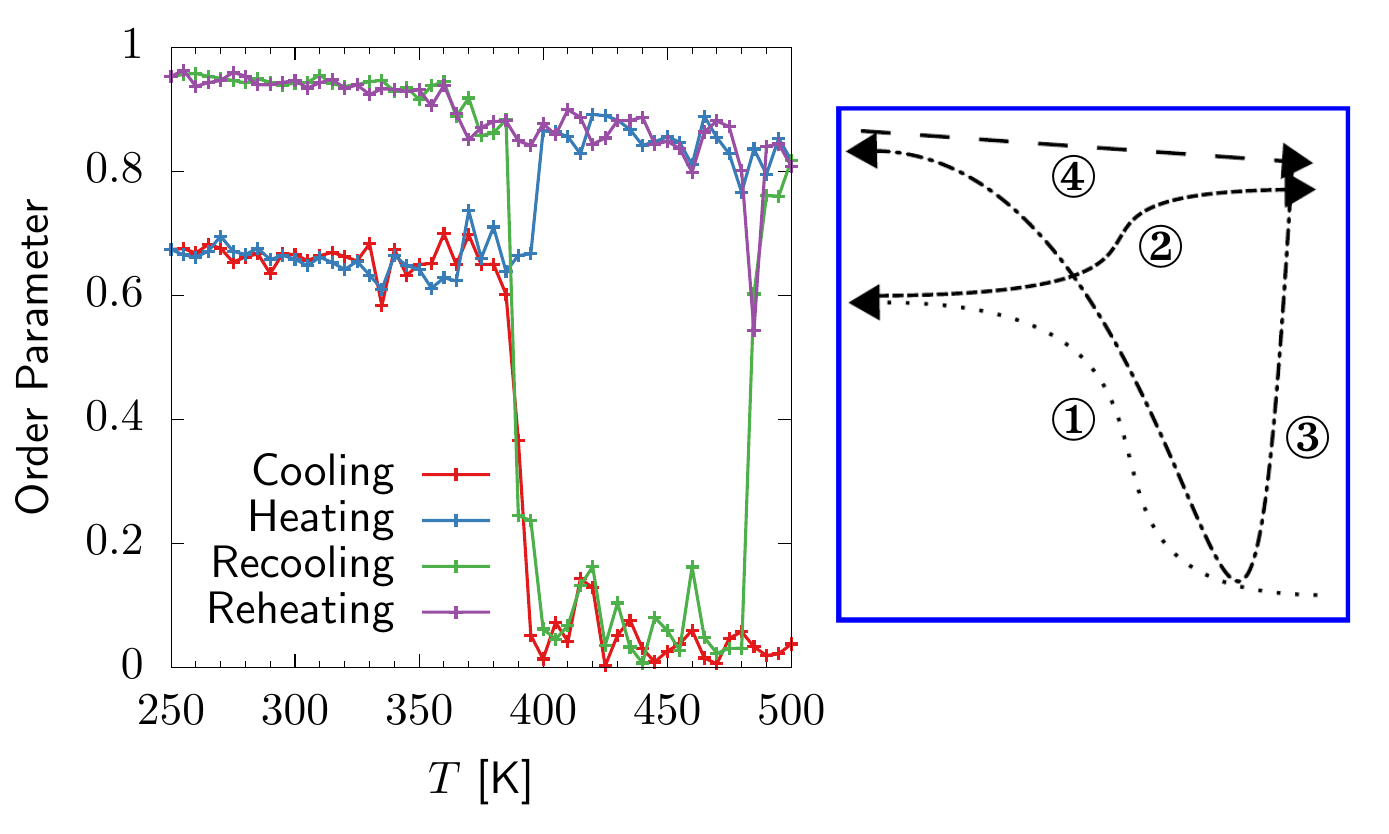}
\caption{Evolution of a representative structural order parameter
  (where a value of 1 is completely ordered and 0 is disordered)
  during the cooling, heating, recooling and reheating CG-MD
  simulations.}
\label{fig:CG-ann}
\end{figure}

Since the CG model of Fritz \emph{et al.}~\cite{Fritz2009} derived
using CRW method from AA simulations can predict melt properties,
including the melt packing and the density between 400 and 520 K, we
use this model to perform annealing simulations to test its ability to
crystallize. The initial system is an sPS melt that contains 9 chains,
each comprising 10 monomers. To crystallize the system, annealing
simulation is performed from $T = 500$~K to $T = 250$~K, with a total
cooling time of $2.5 \times 10^5~\tau$. This is then followed by a
heating procedure back to $T = 500$~K at the same rate. The evolution
of a commonly-used structural order parameter during the cooling,
heating, recooling and reheating processes is displayed in
Fig.~\ref{fig:CG-ann}. Note that the simulation box is rectangular
under isotropic pressure coupling. It shows that highly-ordered
conformations can indeed be obtained from annealing simulations, but
the fast annealing rate leads to significant hysteresis.

\subsubsection{Replica exchange simulations from CG and AA models}

\begin{figure}
\includegraphics[scale=0.7]{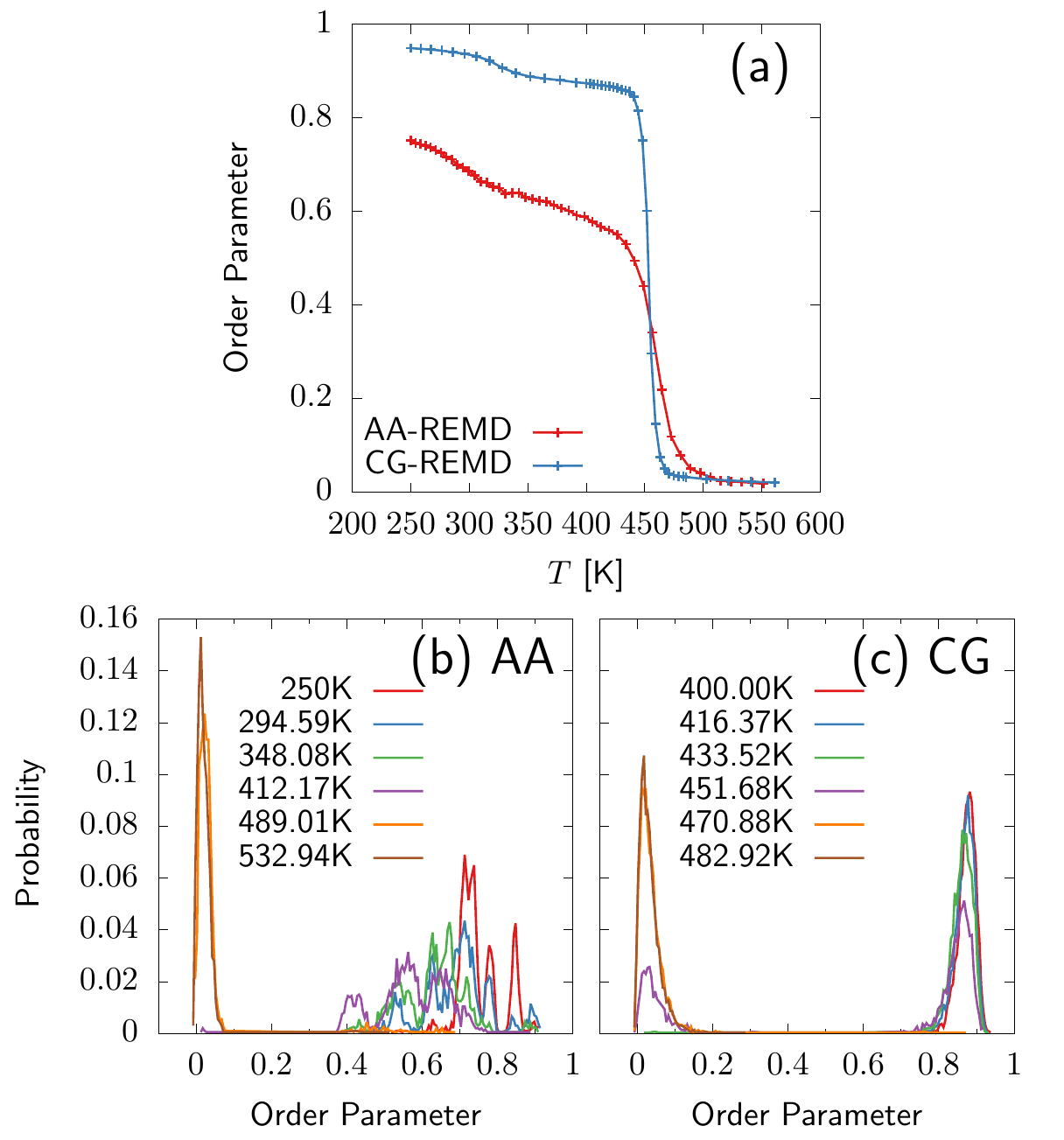}
\caption{(a) Structural order parameter as a function of temperature
  obtained from AA and CG REMD simulations.  Structural
  order-parameter distributions as a function of temperature: (b) AA
  and (c) CG.}
\label{fig:AA-CG-RE}
\end{figure}

To improve the sampling efficiency and accurately characterize the
transition behavior, we employed replica exchange molecular dynamics
(REMD)~\cite{Sugita1999} in the CG and AA simulations. Simulating a
crystallization process using an atomistic model is
difficult. Backmapping is a good strategy to reconstruct an atomistic
configuration from a CG snapshot. We use backmapping to generate a
highly ordered atomistic configuration, which is taken as the initial
structure in the AA simulations. Fig.~\ref{fig:AA-CG-RE}(a) shows that
both AA and CG replica exchange simulations display a phase
transition, and the transition temperatures are both around 450
K. This indicates that this CG model can reproduce the phase
transition of sPS roughly at the correct temperature.  This overall
hints at remarkable transferability of the CG model between the melt
and the crystal.

However, some differences appear between these two models at low
temperatures.  Fig.~\ref{fig:AA-CG-RE} (b) and (c) show multiple peaks
in the order-parameter distributions of the AA model, while only two
peaks are found in the CG model.  The CG model fails to display the
structural heterogeneity of its AA counterpart.

\subsection{Crystal polymorphs}

\begin{figure*}
\includegraphics[scale=1]{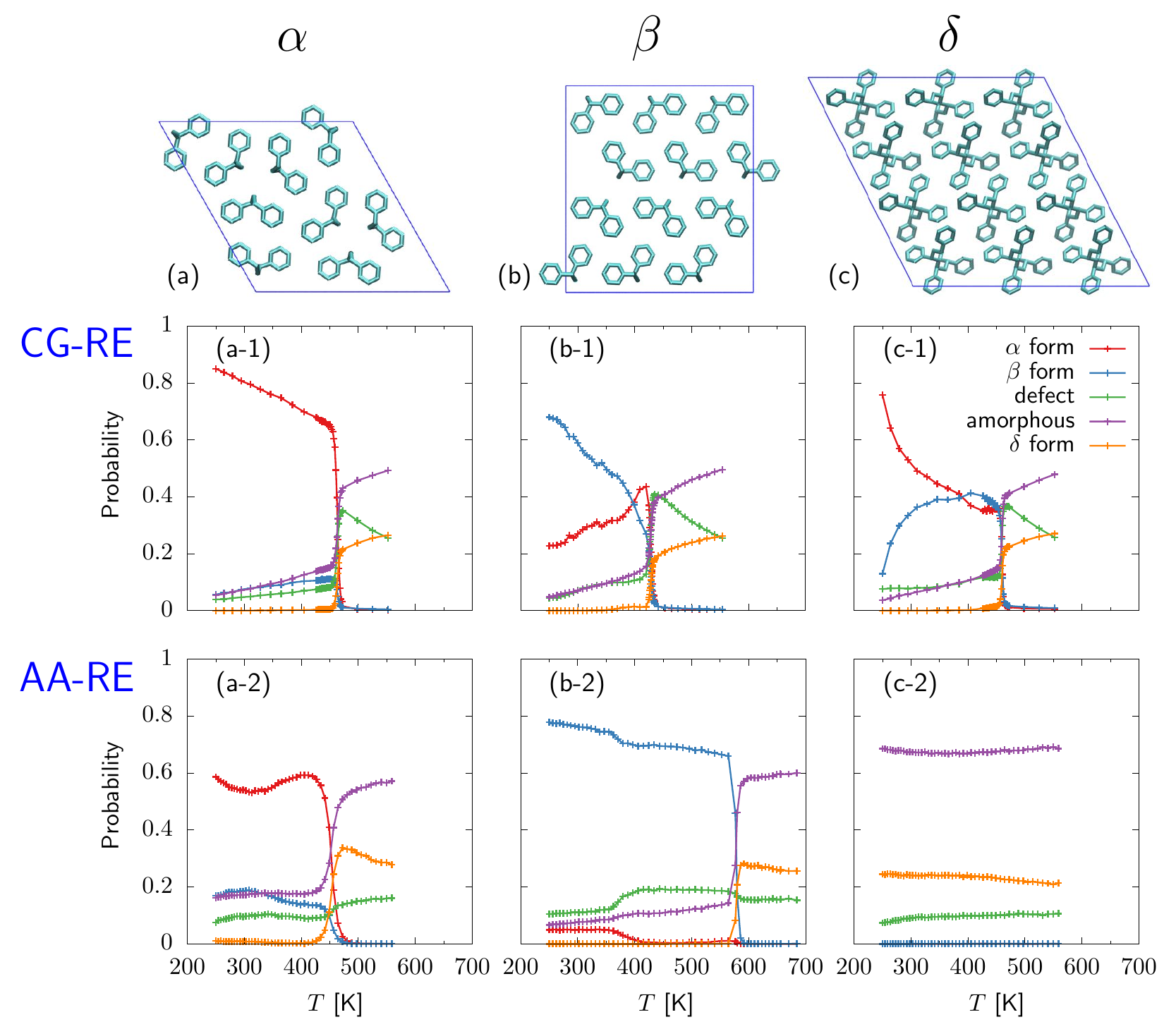}
\caption{From top to bottom: Initial crystal structures for (a)
  $\alpha$, (b) $\beta$ and (c) $\delta$ polymorphs; distributions of
  polymorphs as a function of temperature obtained from CG and AA REMD
  simulations. }
\label{fig:crystal}
\end{figure*}

To further explore crystallization of sPS, we consider the different
crystal forms observed experimentally. The crystalline polymorphs
studied in this paper are listed in Tab.~\ref{tb:Crystal}. We perform
atomistic and coarse-grained replica exchange simulations from these
different crystal structures. Fig.~\ref{fig:crystal} shows that three
different initial configurations lead to different thermodynamics
because of size effects and box shapes.  In the following we analyze
the results in detail.

\begin{table*}
\centering
\caption{ Main crystal structures of s-PS studied in this work, along
  with the number of crystal units, number of chains and number of
  monomers for each chain in the MD simulation.}
\begin{tabular}{l|llllllll}
\hline
 Crystal &  a (\AA) &  b (\AA) &  c (\AA) &  $\gamma$ ($^{\circ}$) &  Space group &  Conformation &  N$_{unit}$ &  $N_{chain}\times N_{mon}$\\
\hline
 $\alpha$~\cite{Rosa1996}   & 25.82 & 26.26 & 5.03 & 119.9 &  $P3$               &  $TTTT$       &  1$\times$1$\times$5 &  9$\times$10\\
 $\beta$~\cite{Chatani1993} & 8.79  & 28.61 & 5.04 & 90.0  &  $P2_{1}2_{1}2_{1}$ &  $TTTT$       &  3$\times$1$\times$5 &  12$\times$10\\
 $\delta$~\cite{Rosa1997}   & 17.38 & 11.73 & 7.81 & 115.0 &  $P2_{1}/a$         &  $(TTGG)_{2}$ &  3$\times$2$\times$3 &  12$\times$12\\
\hline
\end{tabular}
\label{tb:Crystal}
\end{table*}

\subsubsection{$\alpha$ form}

Fig.~\ref{fig:crystal} (a) shows the initial $\alpha$ form
configuration, which contains 9 chains of 10 monomers each.  RE
simulations were performed for both the CG and AA models for
$1.2\times 10^6~\tau$ and 300 ns, respectively.
Fig.~\ref{fig:crystal} (a-1) and (a-2) shows the polymorph
distributions in the temperature range $250-550$~K.  At both
resolutions in this box geometry, the $\alpha$ form is not only stable
but also predominant, with a transition temperature around 460~K.

\begin{figure}
\includegraphics[scale=0.85]{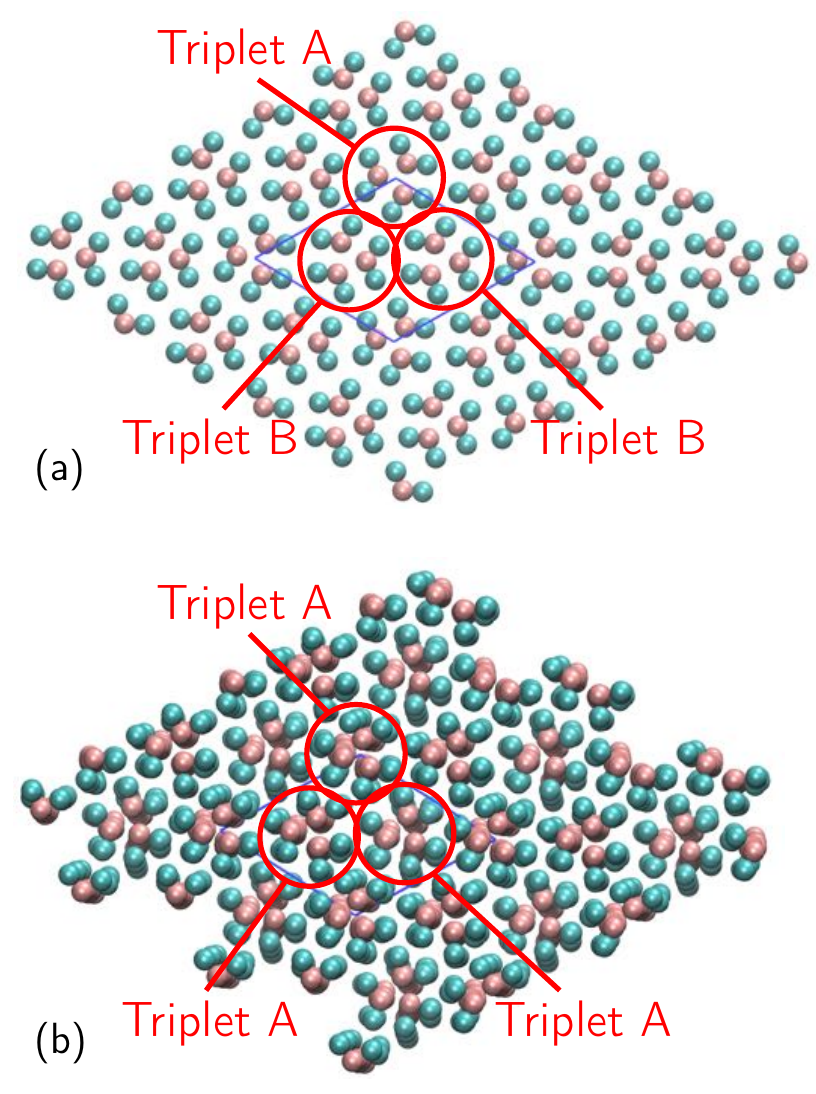}
\caption{(a) The initial $\alpha''$ configuration. (b) A
  representative CG-RE snapshot sampled at 250 K, belonging to the
  $\alpha'$ form.}
\label{fig:CG-alpha}
\end{figure}

Fig.~\ref{fig:CG-alpha} (a) shows the initial CG configuration, set up
in an $\alpha''$ form, as determined by X-ray diffraction
experiments~\cite{Rosa1996}. The limiting-ordered $\alpha''$ form is
characterized by a specific positioning of triplets of chains, e.g.,
one triplet is oriented in one direction, while the other two in the
opposite directions.  The limiting disordered $\alpha'$ form, on the
other hand, has a statistical disorder between these two orientations
of triplets of chains.  Fig.~\ref{fig:CG-alpha} (b) shows a
representative snapshot obtained from CG-RE simulations at 250 K.  We
find that all the triplets of chains sampled from the CG model always
display the same arrangement of triplets---characteristic of
$\alpha'$---while $\alpha''$ does not seem stable in the CG model.

\subsubsection{$\beta$ form}

\begin{figure}
\includegraphics[scale=0.6]{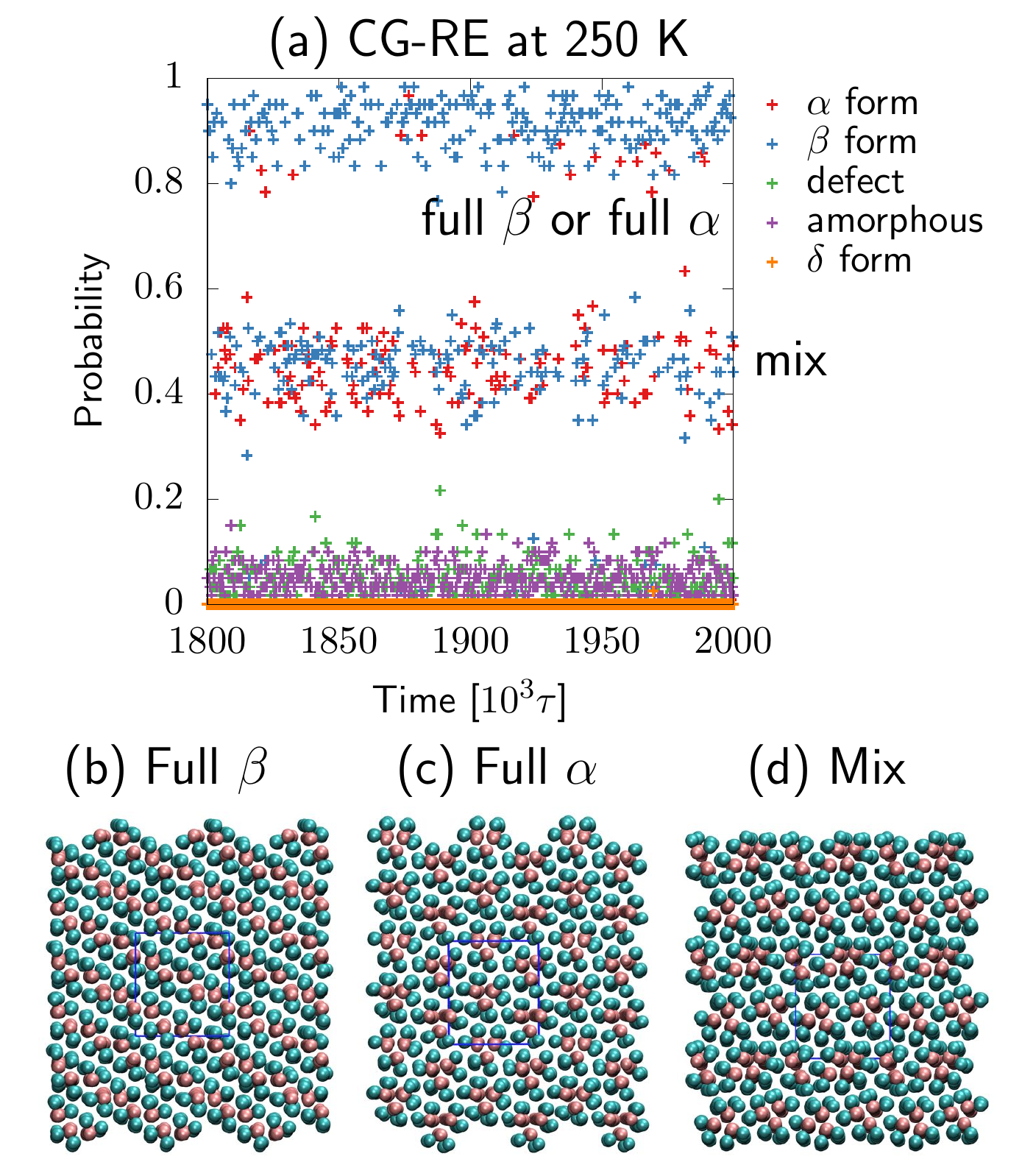}
\caption{(a) Distributions of polymorphs in the last $2\times
  10^5~\tau$ of the CG-RE simulations at 250 K. (b) Full $\beta$, (c)
  full $\alpha$, (d) and mixture at 250K.}
\label{fig:CG-beta}
\end{figure}

Fig.~\ref{fig:crystal} (b) shows the initial $\beta$ configuration,
containing 12 chains of 10 monomers each.  We performed RE simulations
for the CG and AA models for $2\times 10^6~\tau$ and 200~ns,
respectively.  The CG model predominently stabilizes a mixture of
$\alpha$ and $\beta$ forms below the transition temperature (see
Fig.~\ref{fig:crystal} (b-1)).  While many snapshots are made of large
homogeneous regions of $\alpha$ or $\beta$ polymorphs, we also find a
significant population that mix the two, denoted ``mix''
(Fig.~\ref{fig:CG-beta}).  The presence of $\alpha/\beta$ mixtures
breaks the long-range order expected in homogeneous phases and
resembles a form of intermediate between the two polymorphs.  This
mixing has not been reported experimentally.

\begin{figure}
\includegraphics[scale=0.7]{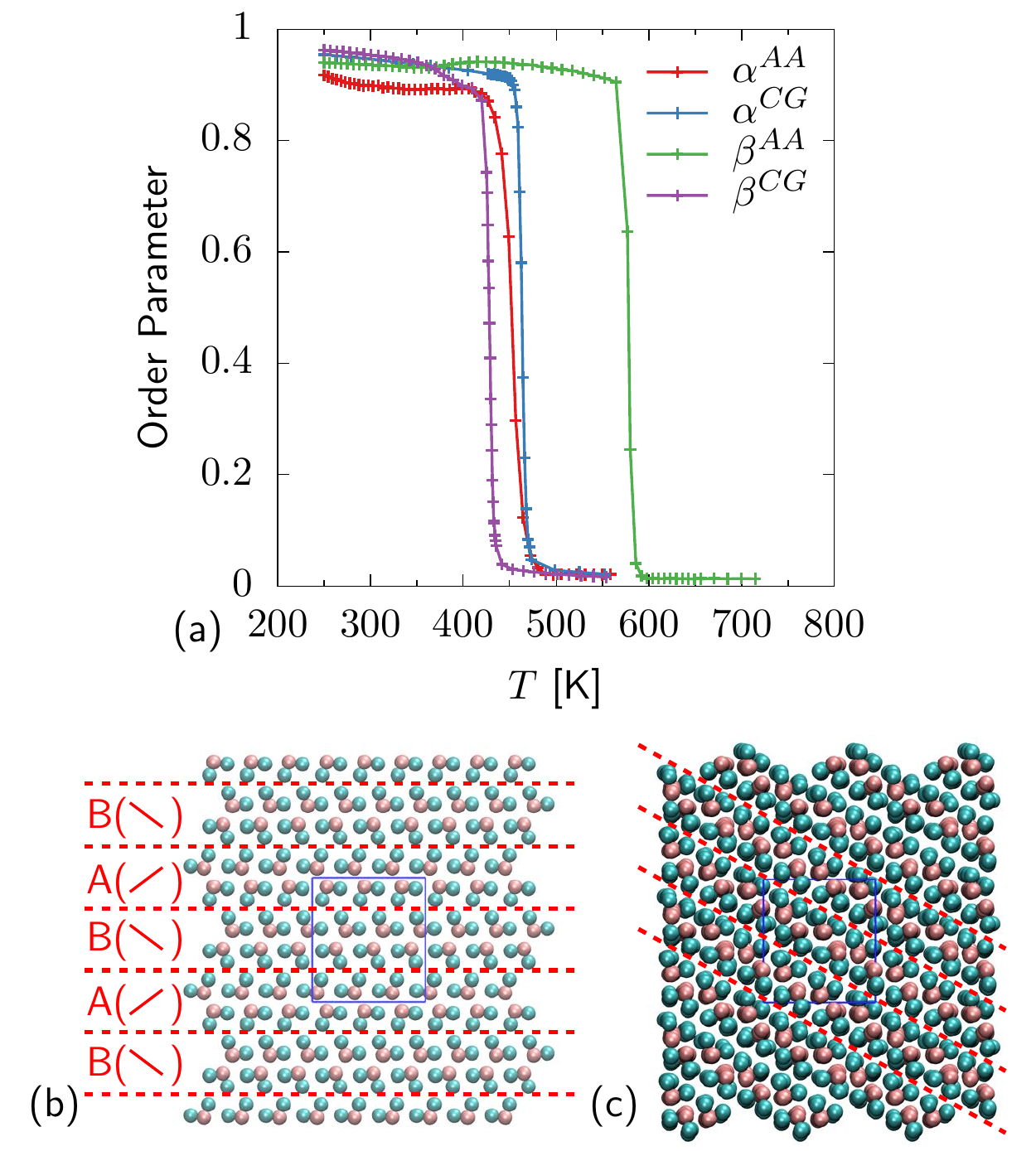}
\caption{(a) Structural order parameter as a function of temperature
  obtained from CG and AA REMD simulations. (b) The initial $\beta''$
  configuration.  (c) Representative snapshot obtained from CG-RE
  simulations at 250~K.}
\label{fig:AA-CG-beta}
\end{figure}

Fig.~\ref{fig:AA-CG-beta}(a) shows the order parameter as a function
of temperature obtained from CG and AA REMD simulations.  Comparing
the two resolutions, we find that the transition temperatures agree
when simulated in the $\alpha$-compatible box shape, but not for
$\beta$.  The AA model shows a much more stable $\beta$ form than the
other scenarios: a transition temperature around 600~K, compared to
$430-450$~K otherwise.  In the CG simulations, the transition
temperature of the $\beta$ form is around 430 K, roughly $20$~K lower
than the $\alpha$ form. Experimentally, crystallization of a melt
leads preferentially to the $\alpha$ ($\beta$) crystals when cooled
from low (high) temperatures~\cite{Woo2001}.  The AA model is thus in
qualitative agreement with the preferential behavior for $\beta$ found
experimentally.

To further study the $\beta$ form in CG simulations, we compare
$\beta$ configurations that are representative of our CG-RE
simulations with the $\beta''$ form observed
experimentally~\cite{Chatani1993}.  Fig.~\ref{fig:AA-CG-beta} (b)
shows there are two kinds of bilayers, indicated as A($\slash$) and
B($\backslash$). The two bilayers are characterized by different
orientations of the lines connecting two adjacent phenyl rings inside
each chain. The regular succession of bilayers ABAB gives rise to the
ordered $\beta''$ form, also showing that these two orientations are
not parallel to the layer lines. On the other hand, CG configurations
have bilayers parallel to the layer lines(Fig.~\ref{fig:AA-CG-beta}
(c)).

The densities obtained from AA and CG simulations are listed in
Tab.~\ref{tb:Density}, compared with experimental data.  While the
temperature dependence shown in Fig.~\ref{fig:AA-CG-beta} (a) resulted
from the average effect of different phases, here we discriminate
between them.  The results show that the AA simulations reproduce the
experimental values satistfactorily.  The CG results are somewhat
larger: While the $\alpha$ form in CG simulations remains in
reasonable agreement with both references, the $\beta$ form is
markedly higher.

These results hint at structural deficiencies for the $\beta$ form in
the CG model, both in terms of limited packing orientations
(\ref{fig:AA-CG-beta}) and higher density.  We hint at the limited
modeling of $\pi$-stacking interactions between side chains, as well
as the packing of phenyl hydrogens (e.g., T-like stacking).  This
structural deficiency, leading to reduced stabilization of the $\beta$
form, may well explain the transition-temperature offset.

\begin{table}
\centering
\caption{ Density of s-PS polymorphs (g/cm$^3$).}
\begin{tabular}{l|lll}
\hline
 \small{Crystal} &  \small{experiment} &  \small{AA simulation} &  \small{CG simulation} \\
\hline
 $\alpha$        & 1.034 & 1.075 & 1.098 \\
 $\beta$         & 1.08  & 1.078 & 1.208  \\
 $\delta$        & 0.977 & - & - \\
\hline
\end{tabular}
\label{tb:Density}
\end{table}

\subsubsection{$\delta$ form}

Fig.~\ref{fig:crystal}(c) shows the initial $\delta$ form
configuration, containing 12 chains with 12 monomers each.  RE
simulations at the CG and AA leve were performed for $1.6\times
10^6~\tau$ and 100 ns, respectively.  Fig.~\ref{fig:crystal} (c-1)
shows that the $\delta$ form is not stable at low temperatures at the
CG level---$\alpha$ and $\beta$ form coexist instead.  This phenomenon
is consistent with experimental studies, since we are not modeling the
cosolvent necessary to stabilize the $\delta$ form.  At 250 K, the
system favors the $\alpha$ form over $\beta$, possibly because the
geometry of the unit cell is more congruent with the former.

At the AA level, we observed no crystallization
(Fig.~\ref{fig:crystal} (c-2)).  While we do expect the system to
eventually crystallize, we suppose that our simulation times--even
using RE---did not allow to probe long-enough timescales to reach any
reasonable equilibrium.  While AA simulations in the $\alpha$ and
$\beta$ geometries did stabilize crystals, we observed little
dynamical exchange among them, further suggesting the sampling
challenge of polymer crystals at an AA resolution.

\subsubsection{Melt structure}

At high temperatures the system consists of amorphous, defects, and a
minor population of $\delta$-type configurations.  Note that
$\delta$-type phases here refer to the presence of helical chain
segments, while defect correspond to chain segments that are planar
but falling out of the $\alpha$ and $\beta$ forms.  A comparison of
the melt populations between the two resolutions shows the larger
occurrence of planar conformations (i.e., defect) at the CG level. On
the other hand, amorphous and helical conformations have a somewhat
higher population at the AA level. This indicates that the CG model of sPS tends to
favor planar arrangements. Here again, the packing of the phenyl
rings may play a role, in accordance with the high density found for
$\beta$.

\subsection{Toward larger systems}

In the previous sections, we demonstrated that the thermodynamic-based
CG model applied here can crystallize and stabilize major polymorphs.
Given the sampling difficulties already met at the AA level, we
consider larger systems only with the CG model: 96 chains of 10
monomers each.  The CG-RE simulations ran for $7\times 10^5~\tau$.

\begin{figure}
\includegraphics[scale=0.65]{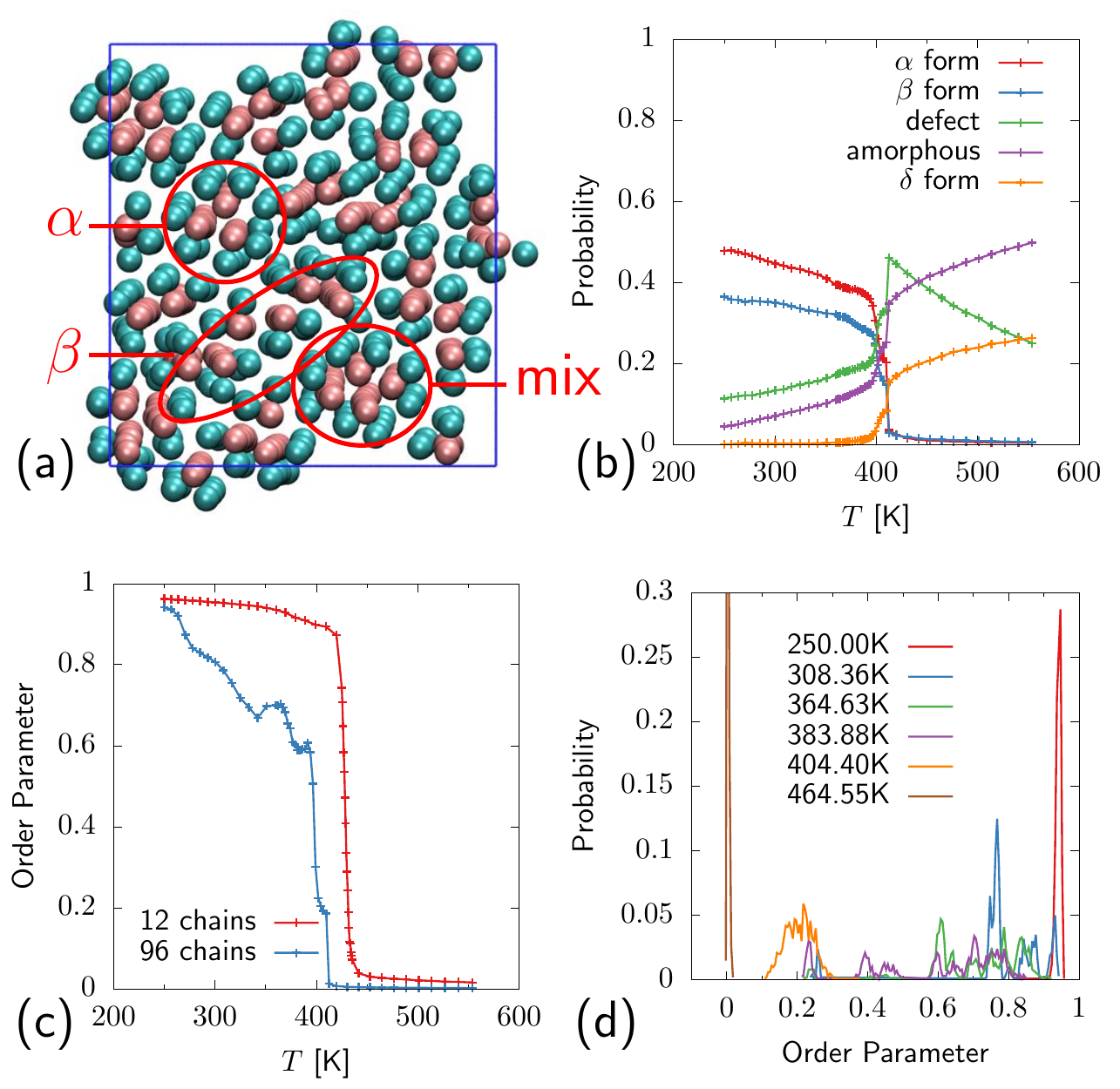}
\caption{CG simulations: (a) Representative snapshot made from 96
  chains at 250K; (b) CG Distributions of polymorphs as a function of
  temperature; (c) Structural-order parameter as a function of
  temperature obtained (comparison between 12 chains and 96 chains);
  and (d) Structural order parameter distributions at different
  temperatures. }
\label{fig:CG-big}
\end{figure}

Fig.~\ref{fig:CG-big} (a) shows one snapshot starting from the $\beta$
form, where $\alpha$, $\beta$ and the mixed phases exist
simultaneously in the crystalline state.  Full $\alpha$ or $\beta$
forms, observed in the small systems, are not found here.  At low
temperatures, the proportion of $\alpha$ is a bit higher than $\beta$
(Fig.~\ref{fig:CG-big} (b)).  It is worth mentioning that we also
performed another simulation starting from the $\alpha$ form (72
chains), and obtained virtually the same proportion of $\alpha$ and
$\beta$ forms.  These converging results thus indicate that they do
not depend on the chosen box geometry, unlike the abovementioned
findings on the smaller systems.  Unfortunately, this result is not
consistent with experiments, which associates a higher stability to
the $\beta$ form.  This stands as another evidence of the structural
deficiency of the $\beta$ form in this CG model.

Fig.~\ref{fig:CG-big}(c) shows that, compared with the small system,
the transition temperature of the large system is a bit lower: 400~K.
Fig.~\ref{fig:CG-big}(d) shows the order-parameter distributions at
different temperatures, which show more structural heterogeneity than
found earlier on (Fig.~\ref{fig:AA-CG-RE} (c)).  We conclude that the
larger system self-assembles more diverse crystalline phases that
dynamically interconvert, leading to an overall reduction in the
transition temperature compared to the small systems.

Interestingly, while the $\alpha$ form was predominantly sampled when
simulating a smaller box congruent with the $\alpha$ unit cell
(Fig.~\ref{fig:crystal}), this preferential stabilization was not
observed here for the larger system.  Experimental results have
suggested that formation of the $\alpha$ form may be a
kinetically-controlled process.\cite{Guerra1990} Our results, while
devoid of non-equilibrium aspects, hint at a templating mechanism that
favors the formation of $\alpha$ due to the environment.


\section{Conclusion}

We report the first computational studies of polymer-crystal
polymorphism.  Replica exchange molecular dynamics simulations are
performed using atomistic (AA) and coarse-grained (CG) models to model
the thermal crystallization of syndiotactic polystyrene (sPS).
Remarkably, the thermodynamic-based CG model not only reproduces melt
properties, but also transfers to the crystals. Furthermore, we find
that the CG model stabilizes the main polymorphs $\alpha$ and $\beta$,
while the $\delta$ form was not observed due to the lack of solvent.
These results are in qualitative agreement with experiments.  Our
simulations suggest the role of templating mechanisms to rationalize
the experimentally-observed kinetic control of the $\alpha$ form.
Because the CG model markedly speeds up the simulations, it can be
used to simulate significantly larger systems, better suited to study
crystallization without finite-size effects.  We thus show that CG
models are powerful tools to investigate the polymorphic behavior of
polymers.  The choices of mapping scheme and force field are important
to stabilize and distinguish the main polymorphs.  In spite of
remaining shortcomings, the ability of the CG model (without
additional tuning) to reproduce the crystallization transition and
polymorphism of polymer is remarkable.  We expect CG simulations to
become a major tool for polymer-crystal polymorphism studies.

\section{Acknowledgment}

We thank Christine Peter, Omar Valsson, and Claudio Perego for
critical discussions and reading of the manuscript.  T.B.~acknowledges
funding from the Emmy Noether program of the Deutsche
Forchungsgemeinschaft (DFG).



\end{document}